\documentclass{document}
 
\usepackage{amsmath}
\usepackage{amssymb}
\usepackage{cite}
\usepackage{cleveref}
\usepackage{dfadobe}  
\usepackage[T1]{fontenc}
\ifpdf
  \usepackage[pdftex]{graphicx} \pdfcompresslevel=9
\else
  \usepackage[dvips]{graphicx}
\fi
\usepackage{pseudocode}
\usepackage{subcaption}

\newcommand{\diff}{\mathrm d}

\title[Real-time High-Quality Rendering of Non-Rotating Black Holes]%
      {Real-time High-Quality Rendering of Non-Rotating Black Holes}

\author[E. Bruneton]{\parbox{\textwidth}{\centering Eric Bruneton}}

\begin{document}


\teaser{
 \vspace{-12pt}
 \includegraphics[width=\linewidth]{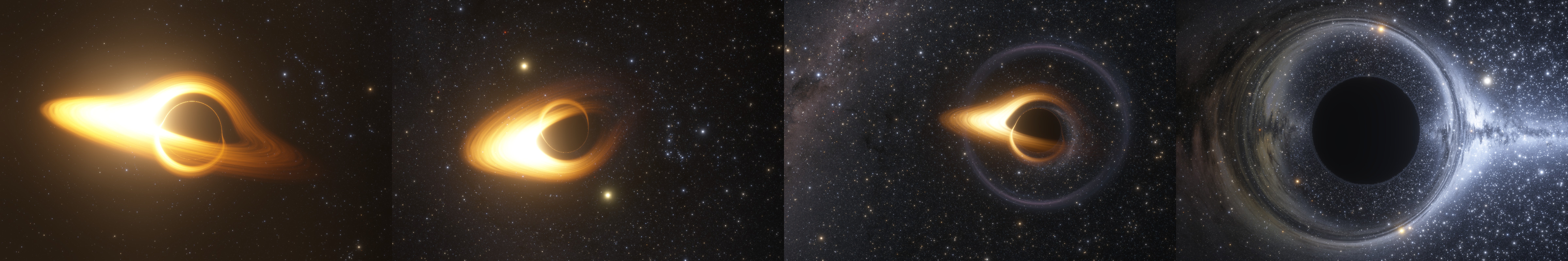}
 \centering
 \caption{Some results obtained with our method. {\em Left}: distorted images 
 of an accretion disc due to gravitational light bending, with relativistic 
 Doppler and beaming effects. {\em Middle}: gravitational lensing creates
 several amplified images of each punctual star and creates Einstein rings. 
 {\em Right}: near the speed of light, light is amplified and blue-shifted 
 ahead, and is reduced and red-shifted behind.
 \vspace{12pt}}
\label{fig:teaser}
}

\maketitle

\begin{abstract}
We propose a real-time method to render high-quality images of a non-rotating 
black hole with an accretion disc and background stars. Our method is based on 
beam tracing, but uses precomputed tables to find the intersections of each
curved light beam with the scene in constant time per pixel. It also uses a 
specific texture filtering scheme to integrate the contribution of the light 
sources to each beam. Our method is simple to implement and achieves high frame 
rates.
\end{abstract}  

\section{Introduction}

Black holes are strange objects which recently got a lot of  public exposure 
with the Interstellar movie~\cite{James2015}, the detection of gravitational 
waves from merging black holes\cite{GRWave2016}, and the first image of a black 
hole~\cite{EHT2019}. A real-time, high-quality visualization of a black hole 
could help the public in getting an intuitive "understanding" of their 
properties, for instance in planetariums or in 3D astronomy software. It could 
also be useful in space games. In this context, we propose a real-time 
high-quality rendering method for non-rotating black holes, with 2 
contributions: a precomputation method for constant time beam tracing, and a 
texture filtering scheme to compute the contribution of the light sources to 
each beam.

We present the related work in Section~\ref{sec:relatedwork}, our model in 
Section~\ref{sec:model} and its implementation in Section~\ref{sec:implem}. We 
conclude with a discussion of our results, limitations and future work in 
Sections~\ref{sec:results} and~\ref{sec:conclusion}.

\section{Related work}\label{sec:relatedwork}

Black hole visualization has a long history starting with~\cite{Luminet1979}, 
and summarized in~\cite{Luminet2019}. Offline rendering methods generally use  
beam tracing in curved space-time, support rotating black holes and produce 
very high-quality images~\cite{Hamilton2014,Riazuelo2014,James2015}. However, 
they are complex to implement and are not interactive ({\em e.g.} an IMAX 
Interstellar frame requires at least 30 minutes with 10 cores and the renderer 
has 40kLoC~\cite{James2015}). Physically accurate general relativistic 
magnetohydrodynamics simulations of accretion discs are even more complex and 
require super-computers~\cite{MNRAS2018}.

\begin{figure*}[htb]
	\centering
	\includegraphics[width=\linewidth]{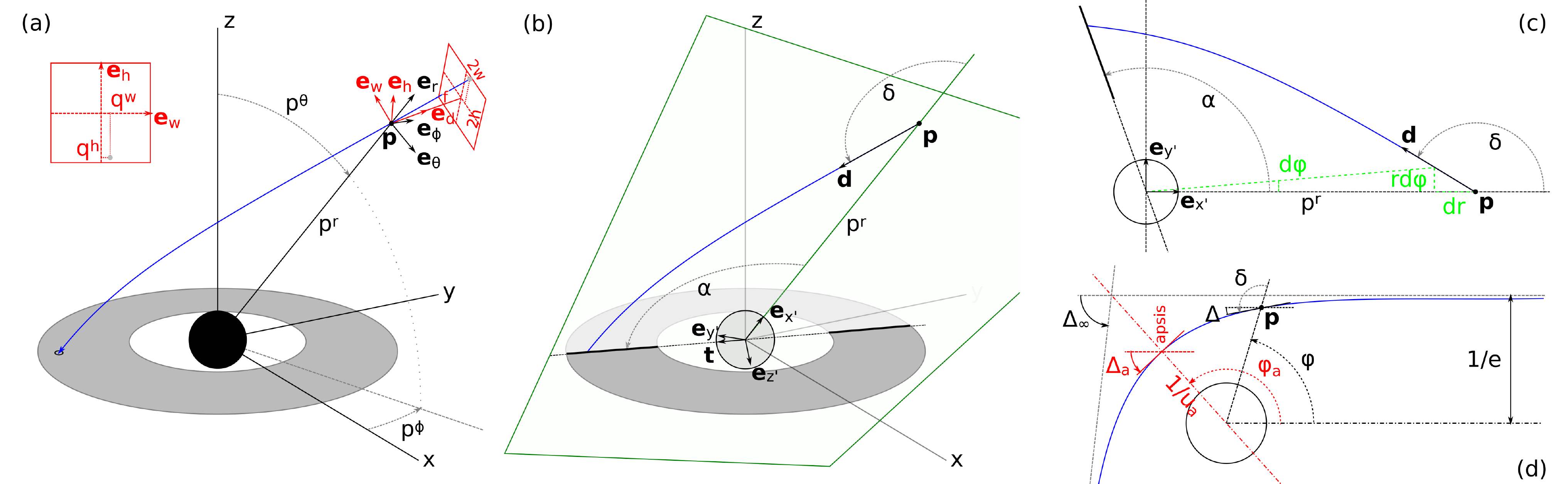}
	{\phantomsubcaption\label{fig:notations:a}}
	{\phantomsubcaption\label{fig:notations:b}}
	{\phantomsubcaption\label{fig:notations:c}}
	{\phantomsubcaption\label{fig:notations:d}}
	\caption{Notations. {\em (\subref{fig:notations:a})} the camera reference 
	frame and image plane (in red) and a curved light ray (in blue) intersecting 
	the accretion disc. {\em (\subref{fig:notations:b})} in the plane containing 
	the light ray, the initial ray angle is noted $\delta$ and the accretion disc 
	inclination $\alpha$. {\em (\subref{fig:notations:c})} $\delta$ verifies 
	$\tan\delta = r \diff\varphi / \diff r$ (in green), {\em i.e.} 
	$\delta = \pi - \arctan2(u, \dot{u})$. {\em (\subref{fig:notations:d})} the 
	deflection $\Delta$ verifies $\Delta = \varphi + \delta - \pi$. The ray is 
	symmetric around the axis through its apsis (in red).}
\end{figure*}

Our work is more related to interactive visualization methods, usually 
restricted to non-rotating black holes to reduce the complexity of the problem.
\cite{Muller2010} render about $120000$ background stars around a black hole, 
whose apparent positions, colors and intensities change due to the 
gravitational effects. Each star is rendered with a point primitive, and its 
projection(s) on screen are found in constant time by using a precomputed 
$4096\times4096$ lookup table. \cite{Muller2011} render a torus and a 
background night sky texture for an observer orbiting a black hole, with a 
ray-tracing method. Ray intersections with the scene are found in constant time 
thanks to lookup tables precomputed with a parallelized code (for a fixed orbit 
radius). \cite{Muller2012} use ray-tracing to render an accretion disc around a 
black hole. Ray intersections are found in constant time by using an analytic 
expression involving the Jacobi-sn function (evaluated with 
arithmetic-geometric, complex number series).

In comparison, our method uses only two small tables ($512\times512$ and 
$64\times32$) which are very fast to precompute. They are used to find, in 
constant time, the intersection(s) of curved light beams with the accretion 
disc and millions of background stars (stored in a cubemap with a specific 
filtering scheme).
  
\section{Model}\label{sec:model}

Our goal is to render a non-rotating black hole with an accretion disc and 
background stars, illustrating the effects of gravitation on light. Simulating 
a realistic accretion disc is {\em not} a goal: we thus use a basic, infinitely 
thin disc model instead. However, we want to get real-time {\em and} 
high-quality images, which is not easy:
\begin{itemize}
	\item a simple ray-marching algorithm can render a sky map texture distorted 
	by a black hole in real-time, but not with a high quality ({\em e.g.} stars 
	become curved segments instead of staying punctual),
	\item conversely, offline beam-tracing methods produce high-quality images 
	but are not real-time~\cite{James2015}.
\end{itemize}
To this end, we propose a "precomputed beam tracing" method: for each pixel, we 
initialize a light beam, compute its intersections with the scene using 
precomputed tables, and then the light received from the intersected objects. 
These 3 steps are explained below, after a very short introduction to the 
Schwarzschild metric.

\subsection{Schwarzschild metric}

The space-time geometry around a non-rotating black hole can be described with 
the Schwarzschild metric. In units such that the radius of the black hole's 
event horizon and the speed of light are 1, this metric is
\begin{equation}
\diff s^2 = \left(1 - \frac{1}{r}\right) \diff t^2 -
            \left(1 - \frac{1}{r}\right)^{-1} \diff r^2 -
            r^2 \diff \theta^2 - 
            r^2 \sin^2\theta \diff \phi^2
\label{eq:metric}
\end{equation}
where $\diff s$ is the line element and $(t, r, \theta, \phi)$ are the 
Schwarzschild coordinates~\cite{weinberg1972}. $r, \theta, \phi$ are 
(pseudo-)spherical coordinates, and in the following we also use the 
corresponding (pseudo-)Cartesian coordinates $x,y,z$, as well as the inverse 
radius $u\triangleq 1/r$. In particular, the Cartesian coordinates of an 
orthonormal basis $\vec{e}_t, \vec{e}_r, \vec{e}_\theta, \vec{e}_\phi$ for 
a static observer at $(t, r, \theta, \phi)$ are, respectively (see 
Fig.~\ref{fig:notations:a})
\begin{equation}
\frac{1}{\sqrt{1-u}}
\begin{bmatrix}
  1\\
  0\\
  0\\
  0
\end{bmatrix},
\sqrt{1-u}
\begin{bmatrix}
  0\\
  \sin\theta \cos\phi\\
  \sin\theta \sin\phi\\
  \cos\theta
\end{bmatrix},
\begin{bmatrix}
  0\\
  \cos\theta \cos\phi\\
  \cos\theta \sin\phi\\
  -\sin\theta
\end{bmatrix},
\begin{bmatrix}
  0\\
  -\sin\phi\\
  \cos\phi\\
  0
\end{bmatrix}
\label{eq:staticbasis}
\end{equation}

\subsection{Beam initialization}

The first step of our method is to compute, for each pixel, the initial 
direction of the corresponding light beam. As \cite{Muller2010}, and in order 
to simplify the next steps, we take advantage of some symmetries to reduce this 
direction to a single angle $\delta$, as shown below.

Let $p=(p^t, p^r, p^\theta, p^\phi)$ be the camera position in Schwarzschild 
coordinates, and $\Lambda$ the Lorentz transformation~\cite{weinberg1972} 
specifying the camera orientation and velocity with respect to a static 
observer at $p$. An orthonormal basis for the camera is thus $\vec{e}_{\tau}, 
\vec{e}_w, \vec{e}_h, \vec{e}_d$ (see Fig.~\ref{fig:notations:a}), given by
\begin{equation}
\vec{e}_i = {\Lambda_i}^j \vec{e}_j, \quad i \in \{\tau, w, h, d\}, \quad j \in 
\{t, r, \theta, \phi\} \label{eq:camerabasis}
\end{equation}
where $\vec{e}_{\tau}$ is the camera 4-velocity and $\vec{e}_w, \vec{e}_h, 
\vec{e}_d$ define its orientation. For a pinhole camera with focal length $f$, 
and since beams are traced backward, the initial beam direction $\mathbf{d}$ 
for a pixel with screen coordinates $q^w, q^h$ is (see 
Fig.~\ref{fig:notations:a})
\begin{equation}
{\mathbf d} = -{\mathbf e}_{\tau} + \frac{q^w {\mathbf e}_w + 
	q^h {\mathbf e}_h - f {\mathbf e}_d}{\sqrt{(q^w)^2 + (q^h)^2 + f^2}}
\label{eq:d}
\end{equation}
where ${\mathbf v}$ denotes the projection of $\vec{v}$ on the $\vec{e}_r, 
\vec{e}_\theta, \vec{e}_\phi$ hyperplane.

We now take advantage of the spherical symmetry of the metric, and of the fact 
that its geodesics are planar~\cite{weinberg1972}, to reduce ${\mathbf d}$ to a 
single angle. Let $(t, r, \vartheta, \varphi)$ be {\em rotated} Schwarzschild 
coordinates such that the beam's axial ray is contained in the equatorial plane 
$\vartheta = \pi/2$. They can be defined as the (pseudo-)spherical coordinates
corresponding to the following new orthonormal basis vectors (for the Euclidean
metric -- see Fig.~\ref{fig:notations:b}):
\begin{equation}
{\mathbf e}_{x'} \triangleq \frac{{\mathbf p}}{p^r} \quad
{\mathbf e}_{y'} \triangleq 
    \frac{{\mathbf e}_{z'} \wedge {\mathbf e}_{x'}}
         {\Vert {\mathbf e}_{z'} \wedge {\mathbf e}_{x'} \Vert} \quad
{\mathbf e}_{z'} \triangleq 
    \frac{{\mathbf e}_{x'} \wedge {\mathbf d}}
         {\Vert {\mathbf e}_{x'} \wedge {\mathbf d} \Vert}
\label{eq:raybasis}
\end{equation}
In these rotated coordinates the metric~\eqref{eq:metric} keeps the same form 
and the light beam starts from $(p^t, p^r, \pi/2, 0)$ with an initial angle 
$\delta \triangleq \arccos({\mathbf e}_{x'} \cdot {\mathbf d} / \Vert {\mathbf 
d} \Vert)$ from the $x'$ axis (see Fig.~\ref{fig:notations:b}).

Finally, note that in the $\vartheta = \pi/2$ plane the accretion disc becomes 
two line segments at angles $\alpha$ and $\alpha + \pi$ from the $x'$ axis, with
\begin{equation}
\alpha = \arccos({\mathbf e}_{x'} \cdot {\mathbf t}) \quad
    {\mathbf t} \triangleq \pm{\mathbf e}_z\wedge{\mathbf e}_{z'} /
        \Vert {\mathbf e}_z \wedge {\mathbf e}_{z'} \Vert
\label{eq:alpha}
\end{equation}
and where the sign is chosen such that ${\mathbf t} \cdot {\mathbf e}_{y'} \ge 
0$ (see Fig.~\ref{fig:notations:b}).

\subsection{Beam tracing}\label{sec:beamtracing}

The second step of our method is to compute the beam intersections with the 
scene, and the light emitted there. For this we first need to determine the 
geodesic followed by the beam's axial ray.

For light rays $\diff s = 0$, and there exist curvilinear coordinates 
$\sigma$, defined up to an affine transform, such that $(1-u)\diff t / \diff 
\sigma$ and $r^2 \sin^2\vartheta \diff \varphi  /\diff \sigma$ are constant 
along the ray~\cite{weinberg1972}. We can thus choose $\sigma$ such that the 
second constant is $1$, leading to
\begin{equation}
(1 - u)\frac{\diff t}{\diff \sigma} = e \quad \mathrm{and} \quad
    \sin^2\vartheta \frac{\diff \varphi}{\diff \sigma} = u^2 
\label{eq:motionconstants}
\end{equation}
where $e$ happens to be the inverse of the ray's impact parameter (see 
Fig.~\ref{fig:notations:d}). By substituting this in \eqref{eq:metric} with 
$\diff s = 0$ and $\vartheta = \pi/2$ we get the geodesic equation
\begin{equation}
\dot{u}^2 \triangleq \left(\frac{\diff u}{\diff \varphi}\right)^2 
    = e^2 - u^2 (1 - u) \quad \Rightarrow \quad
\ddot{u} = \frac{3}{2} u^2 - u\label{eq:rayequation}
\end{equation}

Integrating this numerically at each pixel, with a high precision, would be too
slow (see Section~\ref{sec:results}). Alternatively, the analytic solution for 
$u(\varphi)$, using the Jacobi-sn function (not available on GPU), could be 
implemented with numerical series~\cite{Muller2012}. However, we also need the 
retarded time (to animate the accretion disc) and the light ray deflection (for 
the stars). To compute all this easily and efficiently, we use instead two 
small precomputed tables. We explain below how we precompute and use these 
tables to find the beam intersections, thanks to some ray properties that we 
present first.

\subsubsection{Ray properties}\label{sec:properties}

Light rays can be divided in 3 types. If $e^2$ is larger than the maximum $\mu 
\triangleq 4 / 27$ of $u^2 (1 - u)$ over $[0,1]$, reached at the {\em photon 
sphere} $u = 2 / 3$, then \eqref{eq:rayequation} shows that all values of $u$ 
are possible. The light ray thus comes from infinity into the black hole, or 
vice-versa. Otherwise, some values around $2 / 3$ are excluded. The ray either 
stays in the (empty) region $u > 2 / 3$, or comes from infinity, reaches an 
apsis $u_a < 2 / 3$, and goes back to infinity. In the later case $u_a$ is 
given by setting $\dot{u} = 0$ in~\eqref{eq:rayequation}:
\begin{equation}
u_a=\frac{1}{3} + \frac{2}{3}
\sin\left(\frac{1}{3} \arcsin\left(\frac{2e^2}{\mu} - 1\right)\right)
\label{eq:uapsis}
\end{equation}
and the light ray is unchanged by the reflection $\varphi \rightarrow 
2\varphi_a - \varphi$ (see Fig.~\ref{fig:notations:d}). In any case, $\varphi 
\rightarrow -\varphi$ changes a solution of~\eqref{eq:rayequation} into 
another, with $e, \sigma, \dot{u}$ and $\alpha$ changed into their opposite and 
$\delta$ into $\pi - \delta$.

\begin{algorithm}[htb]
	\centering
	\begin{pseudocode}
		\PROCEDURE{Precompute}{\epsilon}
		\FORALL e \ge 0 \DO
		\BEGIN
		t\GETS 0,\ u \GETS 0,\ \dot{u} \GETS e,\ \varphi \GETS 0,\ \diff\varphi 
		\GETS \epsilon \\
		\WHILE u < 1 \AND (\dot{u} \ge 0 \OR \varphi < \pi) \DO
		\BEGIN
		\IF \dot{u} \ge 0 \THEN
		  {\mathbb D}(e,u) \GETS [t,\ \Delta=\varphi - \mathrm{arctan2}(u, 
		  \dot{u})]\\
		\IF \varphi < \pi \THEN {\mathbb U}(e,\varphi) \GETS [t,\ u]\\
		\IF u > 0 \THEN t \GETS t + e\,\diff\varphi / (u^2 - u^3)\\
		\dot{u}\GETS \dot{u}+(3u^2/2-u)\diff\varphi,\ u \GETS 
		u+\dot{u}\diff\varphi,\ \varphi \GETS 
		\varphi+\diff\varphi
		\END
		\END
		\ENDPROCEDURE
		\\
		\PROCEDURE{TraceRay}{p^r,\delta,\alpha,u_{ic},u_{oc}}
		u \GETS 1/p^r,\ \dot{u} \GETS -u\cot\delta,\ e^2 \GETS \dot{u}^2 + u^2(1 - 
		u)\\
		\IF e^2 < \mu \AND u > 2/3 \THEN \RETURN{\infty,\emptyset}\\
		s \GETS \mathrm{sign}(\dot{u}),\ [t, \Delta] \GETS {\mathbb D}(e,u),\ [t_a, 
		\Delta_a] \GETS {\mathbb D}(e,u_a)\\
		\varphi \GETS \Delta + (s = 1\ ?\ \pi-\delta\ :\ 
		\delta) + s \alpha,\ \varphi_a \GETS \Delta_a + \pi / 2\\
		\varphi_0 \GETS \varphi \mod \pi,\ [t_0,u_0] \GETS {\mathbb U}(e, 
		\varphi_0), I \GETS \emptyset\\
		\IF \varphi_0 < \varphi_a \AND u_{oc} \le u_0 \le u_{ic} \AND 
		\mathrm{sign}(u_0 - u) = s \THEN \DO
		\BEGIN
		I \GETS I \cup [s(t_0-t),\ u_0,\ \alpha + \varphi - \varphi_0]\\
		\END\\
		\IF e^2<\mu \AND s=1 \THEN \DO
		\BEGIN
		\varphi \GETS 2\varphi_a - \varphi,\ \varphi_1 \GETS \varphi \mod \pi,\ 
		[t_1,u_1] \GETS {\mathbb U}(e, \varphi_1)\\
		\IF \varphi_1 < \varphi_a \AND u_{oc} \le u_1 \le u_{ic} \THEN \DO
		\BEGIN
		I \GETS I \cup [2t_a-t-t_1,\ u_1,\ \alpha + \varphi - \varphi_1]\\
		\END\\
		\END\\
		\IF \dot{u}>0 \THEN \Delta \GETS (e^2<\mu\ ?\ 2\Delta_a-\Delta\ :\ \infty)\\
		\RETURN{\delta'=\delta+\Delta,I}\\
		\ENDPROCEDURE
	\end{pseudocode}
	\caption{\label{fig:algo} \textsc{Precompute} is based  
	on~\eqref{eq:motionconstants}, \eqref{eq:rayequation} and the properties 	
	illustrated in Fig.~\ref{fig:notations:d}. \textsc{TraceRay} uses the 	
	properties and symmetries presented in Section~\ref{sec:properties}.}
\end{algorithm}

\subsubsection{Precomputations and beam tracing: background stars}

For background stars we first compute the beam's escape angle, and then sum the 
light emitted by all the stars in the beam's footprint on the celestial sphere, 
around this escape direction.

\paragraph*{Escape angle} Let $\delta'$ be the beam's escape angle (or $\infty$ 
if it falls into the black hole), measured from the $x'$ axis. For efficient 
rendering, $\delta'$ could be precomputed for all initial conditions $p^r, 
\delta$. But this would yield an $O(n^3)$ algorithm. Instead, we precompute the 
deflection $\Delta$ of rays coming from infinity (see 
Fig.~\ref{fig:notations:d}) in a $\mathbb{D}(e, u)$ table, for all $e \ge 0$ 
and $u < 1$ or $u \le u_a$ (depending on $e$ and taking advantage of the above 
symmetries). This gives a trivial $O(n^2)$ algorithm (see 
Algorithm~\ref{fig:algo} -- we use the Euler method but Runge-Kutta or other 
methods are possible too). At runtime, we compute $\delta'$ as $\delta + 
\Delta$ or $\delta + \Delta_{\infty} - \Delta = \delta + 2\Delta_a - \Delta$, 
depending on the ray direction (see Fig.~\ref{fig:notations:d} and 
Algorithm~\ref{fig:algo}).

In practice $\mathbb{D}(e, u)$ is defined only in a subset $\mathcal{D}$ of 
$[0, \infty[ \times [0, 1[$, diverges at $(\sqrt{\mu}^{\,-},u_a)$ and 
$(\sqrt{\mu}^{\,+},1)$, and varies rapidly around $u = 2 / 3$ (because rays 
make more and more turns near the photon sphere before falling or escaping). 
For good precision we thus map $\mathcal{D}$ non-linearly into a square $[0, 
1]^2$ domain, designed to get more samples in these regions (see 
Appendix~\ref{sec:texmapping}).

\begin{figure}[t]
	\centering
	\includegraphics[width=\linewidth]{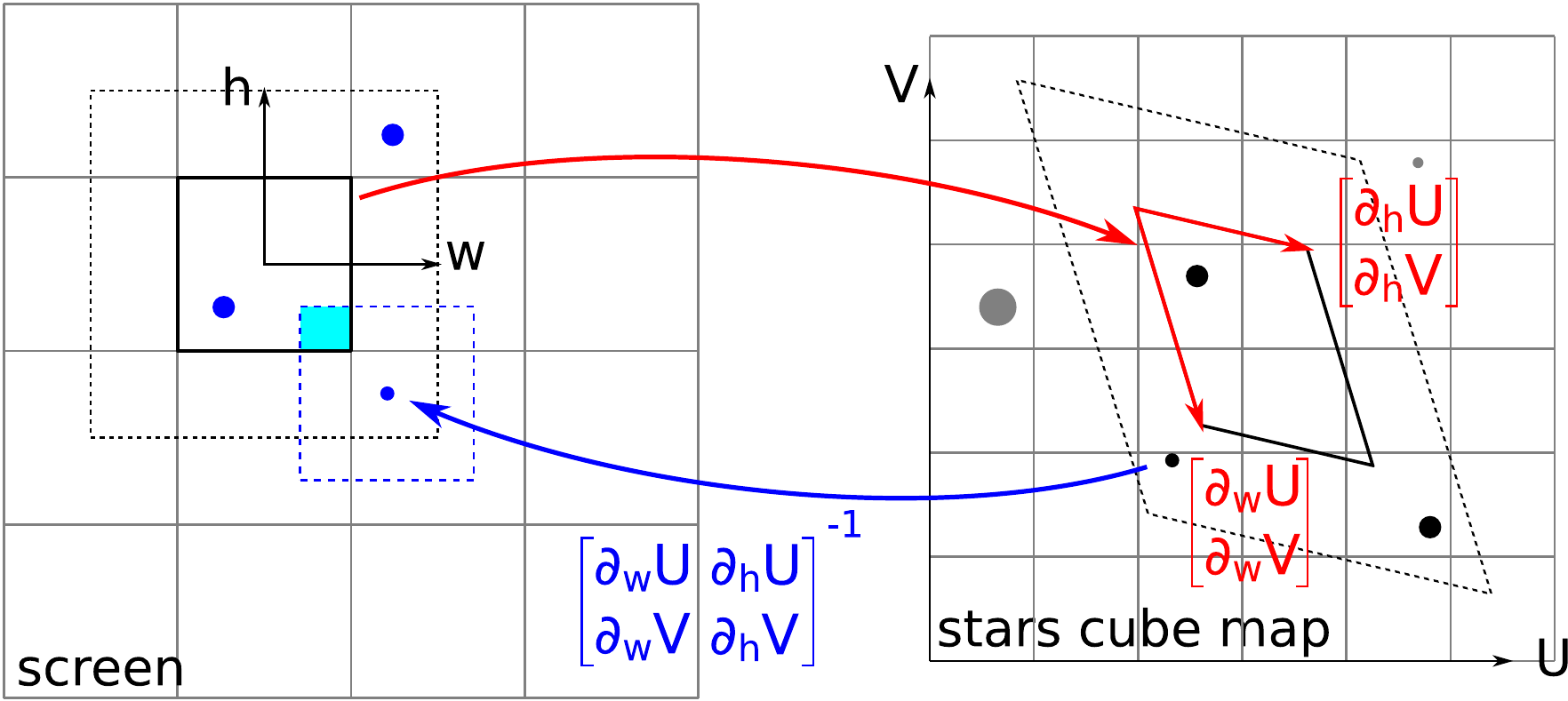}
	\caption{\label{fig:filtering}Filtering. We compute the emitted light for a 
	pixel by summing the light from the stars in its extended footprint (dashed 
	parallelogram, computed with screen space partial derivatives) in a stars 
	map, weighted by their pixel overlap area (cyan).}
\end{figure}

\paragraph*{Emitted light} It now remains to compute the light emitted from all 
the stars in the beam's footprint on the celestial sphere, around $\mathbf{d}' 
= \cos\delta' \mathbf{e}'_x + \sin\delta' \mathbf{e}'_y$. For extended sources 
such as nebulae or galaxies, we can simply take advantage of anisotropic 
texture filtering, by storing these sources in a cube map. For punctual stars, 
however, this would yield unrealistically stretched star images. Our solution 
is to use a manually filtered cube map (see Fig.~\ref{fig:filtering}):
\begin{itemize}
	\item Each texture element (or {\em texel}) stores the color and position 
	(in the texel) of at most one star. A color {\em sum} (and not average) 
	and luminosity-weighted position average is used for mip-mapping.
	\item We compute the beam's footprint in the cube map by using screen space 
	partial derivatives (implemented with finite differences by the rendering 
	pipeline). To avoid discontinuities at cube edges, we compute the partial 
	derivatives $\partial_w \mathbf{d}'$ and $\partial_h \mathbf{d}'$ of 
	$\mathbf{d}'$, and then compute the derivatives of the cube map face texture 
	coordinates $U, V$ analytically from them ({\em e.g.} for the +z face $U = 
	d'^x / d'^z$, $\partial_w U = (\partial_w d'^x - U\partial_w d'^z) / d'^z$).
	\item We compute a mipmap level from the size of the footprint, fetch all the 
	texels at this level in the footprint, and accumulate the colors of the	
	corresponding stars. For anti-aliasing, and to conserve the total intensity, 
	we view each star as a $1\times 1$ area and multiply its intensity with the 
	area of its intersection with the considered screen pixel ({\em i.e.} with 
	$f(w)f(h)$, where $f(x) = \max(1 - |x|, 0)$ and$(w, h)$ is the star's 
	subpixel coordinates -- the pixel domain being $[-\frac{1}{2}, 
	\frac{1}{2}]^2$). Note that this requires to consider an extended footprint 
	(see Fig.~\ref{fig:filtering}). In our implementation, we select the mipmap 
	level so that it is at most $9 \times 9$ texels. 
\end{itemize}
Note that this method approximates quadrilateral footprints with 
parallelograms, and does not use interpolation across mipmap levels. This would 
be easy to fix, but is not really necessary since our method already gives very 
good results.

\subsubsection{Precomputations and beam tracing: accretion disc}

As for stars, we compute the beam intersection(s) with the accretion disc by 
using a precomputed table. We then compute the light emitted there by using a 
simple procedurally animated disc model.

\paragraph*{Intersections} Let $r_{ic}$ and $r_{oc}$ be the inner and outer 
radius of the disc (with $r_{ic}\ge 3$, the innermost stable circular 
orbit~\cite{Lasota2016}). Since the intersections can only occur at 
$\varphi = \alpha + m\pi$ (see Fig.~\ref{fig:notations:b}), we only need the 
function $u(\varphi)$ to check if there exist $m$ such that $u_{oc} \triangleq 
r_{oc}^{-1} \le u(\alpha + m\pi) \le u_{ic} \triangleq r_{ic}^{-1} \le 1 / 3$. 
For this we precompute $u(e, \varphi)$, {\em for light rays coming from 
infinity}, in a $\mathbb{U}(e, \varphi)$ table. At runtime, $u(\alpha + m\pi)$ 
can then be computed with $\mathbb{U}(e, \varphi_p + \alpha + m\pi)$, where 
$\varphi_p$ is the camera position (which can be obtained from the deflection 
$\Delta_p = \mathbb{D}(e, 1/p^r)$ since $\Delta = \varphi + \delta - \pi$ -- 
see Fig.~\ref{fig:notations:d}).

Note that we don't need $\mathbb{U}$ for all $\varphi$: we can stop when $u \ge 
1 / 3$ since no intersection can occur between this point and the apsis, if any 
(and the rest can be deduced by symmetry). In practice, this means that we only 
need $\mathbb{U}(e, \varphi)$ for $0 \le \varphi < \pi$, which has two 
consequences:
\begin{itemize}
	\item we don't need to evaluate $\mathbb{U}(e, \varphi_p + \alpha + m\pi)$ 
	for all $m$: in fact we only need $\mathbb{U}(e, (\varphi_p+\alpha)\mod\pi)$,
	\item there can be at most two intersections: one on each symmetric part of 
	the ray (if it does not fall into the black hole).
\end{itemize}
The algorithms to precompute $\mathbb{U}(e, \varphi)$ and to find the accretion 
disc intersections $(u_0, \varphi_0)$, $(u_1, \varphi_1)$ follow from these 
properties, and those of Section~\ref{sec:properties}, and are shown in 
Algorithm~\ref{fig:algo}. As for $\mathbb{D}$, we map $\mathbb{U}$'s domain 
non-linearly into $[0,1]^2$ to get good precision in large gradient areas (see 
Appendix~\ref{sec:texmapping}).

Finally, note that for an animated disc we also need to compute the {\em 
retarded time} between the intersections and the camera. For this we also 
precompute and store $t$ -- using $\diff t / \diff\varphi = e / (u^2 - u^3)$, 
from \eqref{eq:motionconstants} -- in $\mathbb{D}$ and $\mathbb{U}$. This 
allows the computation of the retarted times $t_0$, $t_1$ at the disc 
intersections, as shown in Algorithm~\ref{fig:algo}.

\begin{figure}[t]
	\centering
	\includegraphics[width=\linewidth]{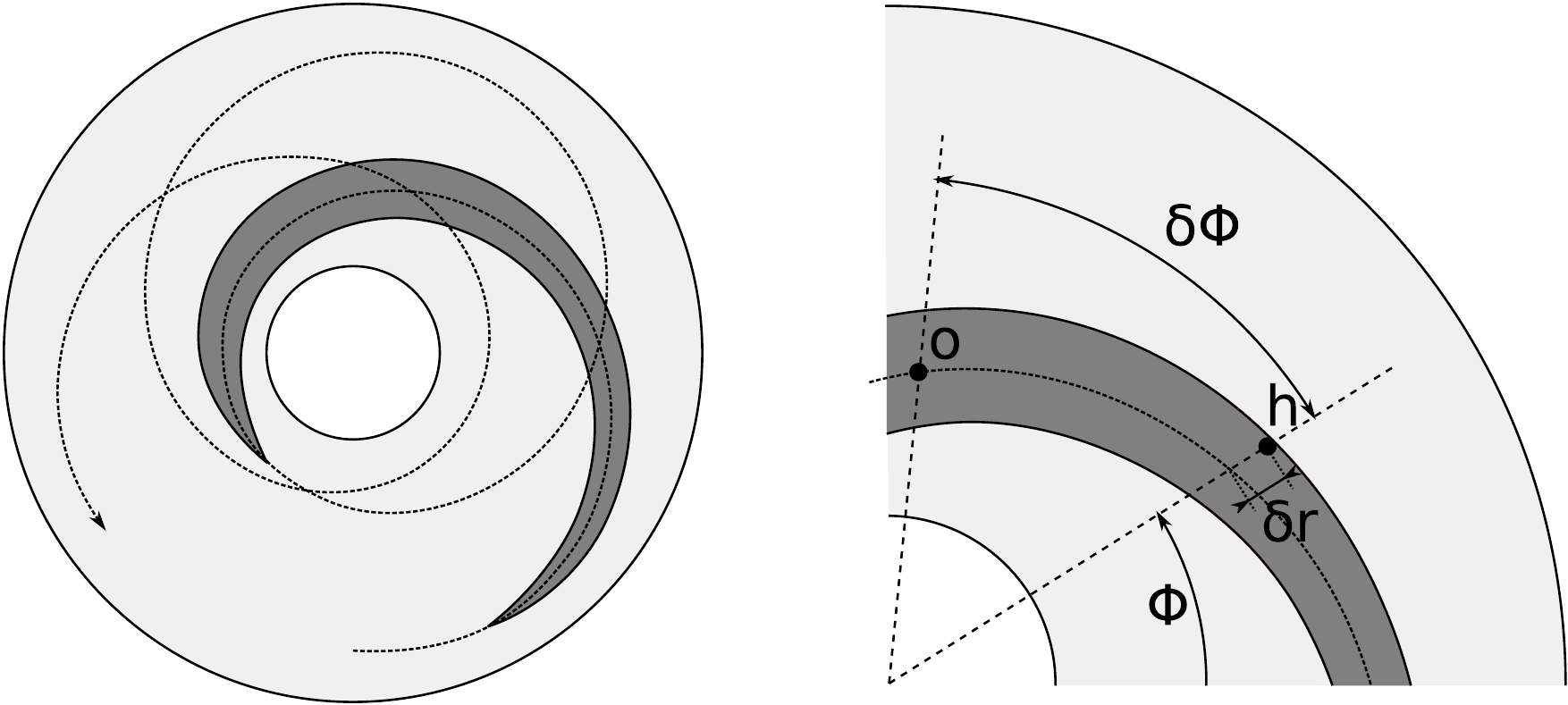}
	\caption{\label{fig:disc}Accretion disc. We compute the density with a 
	sum of linear particles moving along precessing orbits (left), whose density 
	at a point $h$ depends on its "distance" $d(\delta r, \delta\phi)$ from the 
	particle center $o$ (see Appendix~\ref{sec:discparticles}).}
\end{figure}

\paragraph*{Emitted light} It now remains to compute the light emitted by the 
accretion disc at the intersection points. For this we use the light emitted by 
a black body at temperature $T(u)$, times the disc density. We use $T^4(u) 
\propto u^3 (1 - \sqrt{3u})$~\cite{Lasota2016}, and compute the density with a 
sum of procedural particles moving along quasi-circular precessing orbits (see 
Fig.~\ref{fig:disc} and Appendix~\ref{sec:discparticles}).

\subsection{Shading}

Due to gravitational effects, the light received at the camera is different 
from the emitted light, computed above. We present these effects below, and 
explain how we compute them.

\subsubsection{Gravitational lensing effects}

Due to gravitational lensing, the light emitted by a punctual star is received 
amplified by a factor $\Omega / \Omega'$, where $\Omega$ (resp. $\Omega'$) is 
the beam's solid angle at the camera (resp. emitter)~\cite{Virbhadra99}. We 
compute it by using the screen space partial derivatives of the beam directions 
at the camera and at the emitter:
\begin{equation}
\frac{\Omega}{\Omega'} = 
    \frac{\Vert \partial_w \mathbf{q} \wedge \partial_h \mathbf{q} \Vert}
         {\Vert \partial_w \mathbf{d}' \wedge \partial_h \mathbf{d}' \Vert}
\end{equation}
where $\mathbf{q}$ is the normalized $[q^w, q^h, -f]^\top$ vector. Note that 
this does not apply to area light sources, because the beam's subtended area 
varies in inverse proportion.

\subsubsection{Doppler and beaming effects}

Due to gravitational and relativistic time dilation and length contraction 
effects, the frequency $\nu$ of the received light differs from the emitted 
frequency $\nu'$. The ratio is given by~\cite{Philipp2017}
\begin{equation}
\frac{\nu}{\nu'} = \frac{g(\vec{k}, \vec{l})}{g(\vec{k}', \vec{l}')}
\label{eq:Doppler}
\end{equation}
where $\vec{k}$ is the 4-velocity of the receiver, $\vec{l}$ is the tangent 
4-vector $\diff s / \diff 
\sigma$,  at the receiver, of the light ray curve $s(\sigma)$, and  
$\vec{k}'$ and $\vec{l}'$ are the corresponding emitter quantities. We thus 
need $\vec{k}$, $\vec{k}'$, $\vec{l}$, and $\vec{l}'$, that we compute as 
follows.

In {\em rotated} Schwarzschild coordinates, $\vec{l}$ and $\vec{l}'$ are 
given by $[\diff t / \diff \sigma, \diff r / \diff \sigma, \diff \vartheta / 
\diff \sigma, \diff\varphi / \diff \sigma]^{\top}$. Using 
\eqref{eq:motionconstants}, this gives
\begin{equation}
\vec{l} = \left[ \frac{e}{1 - u}, -\dot{u}, 0, u^2 \right]^{\top} \quad
\vec{l}' = \left[\frac{e}{1 - u'}, -\dot{u'}, 0, u'^2 \right]^{\top}
\end{equation}
where $e$ is the {\em negative} root of \eqref{eq:rayequation} -- for the  
actual light rays $\diff \varphi / \diff t$ is negative, unlike in the previous 
section where rays were traced backward. In {\em non-rotated} Schwarzschild 
coordinates, we have
\begin{equation}
\vec{k}' = \left[
    \sqrt{\frac{2}{2 - 3u'}}, 
    0,
    0,
    \sqrt{\frac{u'^3}{2 - 3u'}} \right]^{\top}
\label{eq:k_prime}
\end{equation}
for the accretion disc (if we assume a circular motion)~\cite{Philipp2017}, 
$\vec{k}' = [1, 0, 0, 0]^{\top}$ for static stars, and $\vec{k} = 
\vec{e}_{\tau}$ for the camera. Finally, to compute $g(\vec{k}, \vec{l})$ and 
$g(\vec{k}', \vec{l}')$, we need the corresponding {\em rotated} coordinates: 
$k^t$ and $k^r$ are unchanged, $k^\vartheta$ is not needed since $l^\vartheta = 
0$ and $k^\varphi = \vec{k} \cdot \vec{\partial}_\varphi/r^2 = u\mathbf{k} 
\cdot \mathbf{e}_{y'}$ (and similarly for $\vec{k}'$). For instance, for the 
accretion disc, we get $k'^\varphi = k'^\phi \mathbf{e}_z \cdot 
\mathbf{e}_{z'}$ and
\begin{equation}
g(\vec{k}', \vec{l}') = e \sqrt{\frac{2}{2 - 3u'}} - 
    \sqrt{\frac{u'^3}{2 - 3u'}} {\mathbf e}_z \cdot {\mathbf e}_{z'}
\end{equation}

The above Doppler effect has an associated {\em beaming} effect: the received 
intensity differs from the emitted one because, from Liouville's theorem, 
$I(\nu) / \nu^3$ is invariant \cite{Misner1973}. In terms of wavelength  
$\lambda \triangleq \nu^{-1}$, and with $I(\lambda) \diff \lambda = I(\nu) 
\diff \nu$, this gives $I(\lambda) = (\lambda' / \lambda)^5 I(\lambda')$. For 
black bodies the two effects result in a temperature shift $T = (\nu / \nu') 
T'$. For other light sources however, that we want to support, the result is 
more complex. We thus precompute it in a 3D texture $\mathbb{C}(xy, D)$, for 
each chromaticity $xy$ and Doppler factor $D \triangleq \nu / \nu'$.

To this end we need to choose a spectrum for each chromaticity, among the 
infinite number of possible spectrums. For simplicity and to get black body 
spectrums for black body colors, we use spectrums of the form $I(\lambda') = 
B_T(\lambda')(1 - a_1 A_1(\lambda') -a_2 A_2(\lambda'))$, where $B_T$ is the 
black body spectrum for temperature $T$, $T$ is the correlated color 
temperature, and $A_1$ and $A_2$ are two fixed absorption spectrums. A linear 
system gives $a_1$ and $a_2$ from the $xy$ chromaticity, and the Doppler and 
beaming effects give CIE XYZ colors that we precompute with
\begin{equation}
\mathbb{C}(xy,D) = D^5
    \frac{\int I(D\lambda) \left[ \bar{x}(\lambda), \bar{y}(\lambda),  
    	        \bar{z}(\lambda)\right]^{\top} \diff \lambda}
         {\int I(\lambda) (\bar{x}(\lambda) + \bar{y}(\lambda) +   
         	    \bar{z}(\lambda))\ \diff\lambda}
\end{equation}
where $\bar{x}$, $\bar{y}$ and $\bar{z}$ are the CIE color matching functions. 
At runtime, the emitted XYZ color computed in Section \ref{sec:beamtracing} is 
transformed into the received color $(X + Y + Z)\mathbb{C}(xy, \nu/\nu')$ 
with \eqref{eq:Doppler}.

\subsubsection{Lens glare effects}

Due to light scattering and diffraction inside the eye, haloes appear around 
very bright light sources, which would otherwise be hard to distinguish from 
fainter sources. For this reason we apply a bloom shader effect on the final 
image, before tone-mapping. We use a series of small support filter kernels on 
mipmaps of the full image, approximating a point spread function from 
\cite{Spencer95}, but more precise methods are possible too 
\cite{HullinSIG2011}.

\section{Implementation}\label{sec:implem}

\begin{figure}[t]
	\centering
	\includegraphics[width=\linewidth]{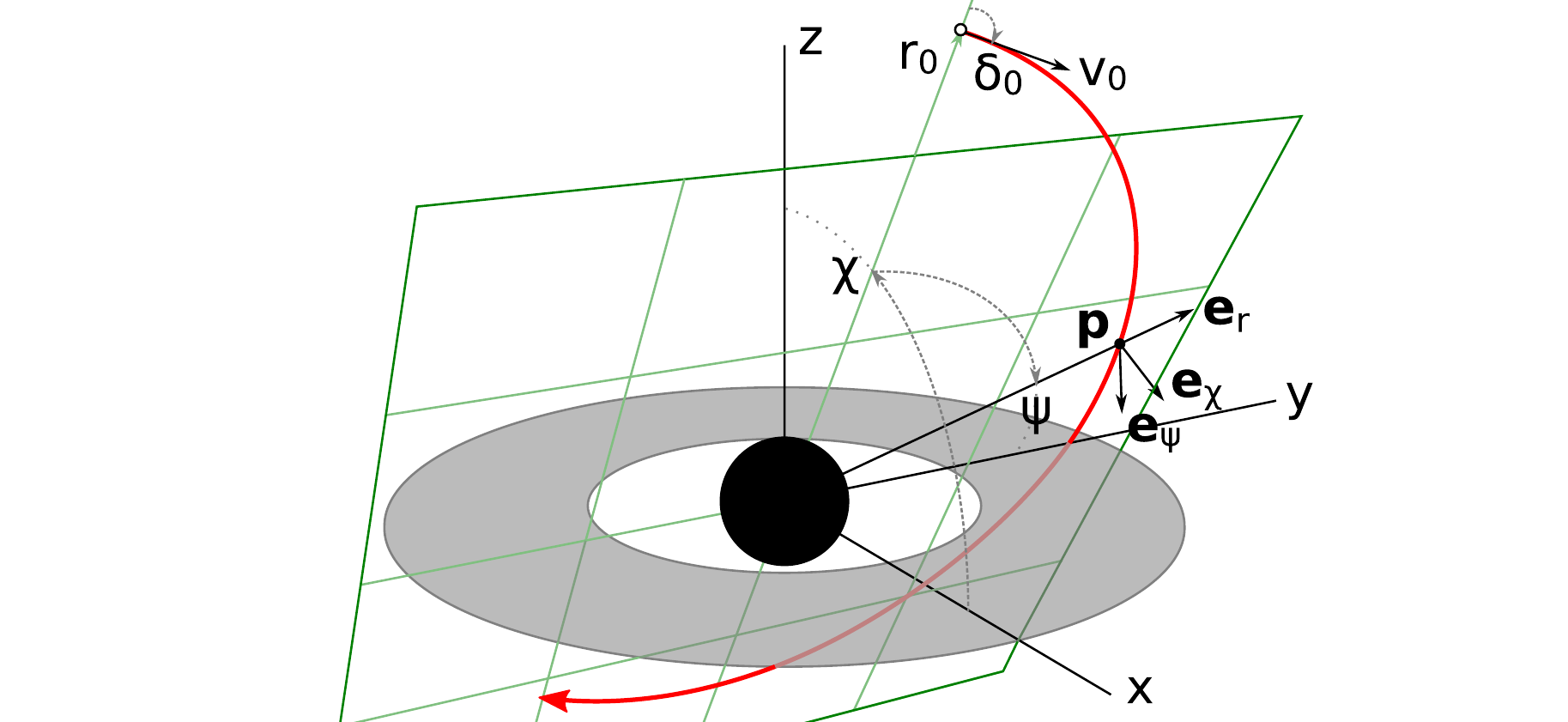}
	\caption{\label{fig:orbit}Camera orbit. The orbit, in red, is specified by 
	an inclination $\chi$ and the initial conditions $r_0$, $\delta_0$ and $v_0$ 
	(see Appendix~\ref{sec:cameraorbit}).}
\end{figure}

We implemented our method in C++ for the precomputations, and WebGL 2 for the 
rendering. The full source code and an online demo are available at 
\url{https://github.com/ebruneton/black_hole_shader}. The demo simulates a 
static or freely falling observer (see Fig.~\ref{fig:orbit}) and allows the 
user to set various parameters (black hole mass, disc temperature and density, 
camera orbit, etc).

We precompute $\mathbb{D}(e, u)$ and $\mathbb{U}(e, \varphi)$ in $512 \times 
512$ and $64 \times 32$ RG32F textures, respectively. \textsc{Precompute} takes 
about 11 seconds on a $3.2$GHz Intel Core i5-6500 CPU, with $\epsilon = 
10^{-5}$, and unit tests show that \textsc{TraceRay} results $\delta'$, $u_0$ 
and $u_1$ are within $10^{-3}$ of reference values computed without 
intermediate textures. We precompute $\mathbb{C}(xy, D)$ in a $64 \times 32 
\times 64$ RGB32F texture, in about 7 seconds.

We also precompute two $6 \times 2048 \times 2048$ RGB9E5 cubemaps for the area 
and punctual light sources, from the Gaia DR2 \cite{gaiadr22018} and Tycho 2
\cite{tycho2000} star catalogs. This requires downloading and processing 550GB 
of compressed data, which can take a day, and yields $\approx 3.6$ million 
punctual light sources. In our implementation we don't store a sub-texel
position for each star: instead, we use a hash of its color to compute a 
pseudo-random position.

\section{Results and discussion}\label{sec:results}

Some results obtained with our method are shown in Fig.~\ref{fig:teaser}. They 
are rendered in Full HD at about $150$ fps on an NVidia GeForce GTX 960 (see 
Table~\ref{table:perf}).

The benefit of our precomputed tables can be measured by replacing 
\textsc{TraceRay} with a ray-marching method integrating \eqref{eq:rayequation} 
numerically (and keeping everything else unchanged). To get the same 
performance as with the precomputed tables, only $25$ integration steps at most 
can be used, and stars end up at several degrees from their correct positions. 
To get (almost) the same precision, up to 1000 integration steps must be used, 
and the framerate drops to about $45$ fps (see method M3 in 
Table~\ref{table:perf}).

The benefit of our custom texture filtering method to render the stars can be 
measured by replacing it with the method from \cite{Muller2010} (and keeping 
everything else unchanged). Each of the $3.6$ million stars is then 
rendered with $n$ $2 \times 2$ anti-aliased point primitives (we used $n = 2$ 
in our tests). The point position is computed with a lookup in a $4096 \times 
4096$ precomputed texture. This gives about $65$ fps (see method M2 in 
Table~\ref{table:perf}). The main bottleneck is due to the fact that, inside 
the region of Einstein rings, many stars project into the same pixel, leading 
to a lot of overdraw.

\begin{table}
	\centering
	\begin{tabular}{|c|c|c|c|c|c|c|}
		\hline
		View & M1 & M2 & M3 & Stars & Disc & Bloom \\
		\hline
		\hline
		1 & 150 & 64 & 43 & 2.99 & 0.82 & 1.11 \\
		2 & 160 & 59 & 45 & 2.64 & 0.78 & 1.15 \\
		3 & 153 & 67 & 45 & 2.99 & 0.69 & 1.10 \\
		4 & 114 & 64 & 41 & 4.14 & 1.87 & 1.08 \\
		\hline
	\end{tabular}
	\caption{\label{table:perf} The framerate obtained with our method (M1), with 
	stars rendered with \cite{Muller2010} (M2), and with \textsc{TraceRay} 
	replaced with ray-marching (M3), on the views shown in Fig.\ref{fig:teaser} 
	and numbered from left to right ($1920\times1080$p, NVidia GeForce GTX 960). 
	The other columns give the time used per frame (in milliseconds), to render 
	the stars, the accretion disc and the bloom effect with our method.}
\end{table}

A limitation of our method is that views from inside the horizon are not 
supported. Indeed, since no observer can remain static in this region, we can 
no longer specify the camera and initialize light beams by using a reference 
static observer. Also the Schwarzschild metric diverges at the horizon. 
However, by using different coordinates, as in \cite{Muller2010}, we believe 
that our method can be extended to support this case.

Another limitation is that motion blur, which is necessary for very high 
quality animations, is not supported. Also, because of the approximations in 
our custom texture filtering method, in some cases a few stars flicker when the 
camera moves. Fixing both quality issues might be easier by extending 
\cite{Muller2010} rather than by extending our method, at the price of 
decreased performance.

Finally, another limitation of our method, and of \cite{Muller2010} as well, is 
that rotating black holes are not supported. Because they are only axially 
symmetric, the "inverse ray tracing" approach of \cite{Muller2010} would 
probably be hard to generalize to this case. Our precomputed ray tracing 
method, on the other hand, could in principle be generalized to 4D tables 
containing the deflected direction and the accretion disc intersections of any 
ray (specified with 2 position and 2 direction parameters). In practice 
however, obtaining precise 4D tables of reasonable size might be hard.

\section{Conclusion}\label{sec:conclusion}

We have presented a beam tracing method relying on small precomputed textures 
to render real-time high-quality images of non rotating black holes. Our method 
is simple to implement and achieves high frame rates. Extending it to views 
from inside the horizon, and to rotating black holes, if this is possible, is 
left as future work.

\paragraph*{Acknowledgments} We would like to thank Alain Riazuelo for 
proofreading this paper.

\appendix
\section{Texture mappings}\label{sec:texmapping}

We store $\mathbb{D}(e,u)$ at texel coordinates
\begin{equation*}
\begin{split}
&\left[\frac{1}{2} - \sqrt{-\log(1 - e^2 / \mu) / 50},\ 1 - \sqrt{1 - u / 
u_a}\right]\ \mathrm{if}\ e^2 < \mu\\
&\left[\frac{1}{2} + \sqrt{-\log(1 - \mu / e^2) / 50},\ \frac{\sqrt{2/3} \pm 
\sqrt{|u - 2 / 3|}}{\sqrt{2 / 3} + \sqrt{1 / 3}}\right]\ \mathrm{otherwise}
\end{split}
\end{equation*}
where $\pm$ is the sign of $u-2/3$, and $\mathbb{U}(e, \varphi)$ at texel 
coordinates
\begin{equation*}
\left[\frac{1}{1 + 6 e^2},\ \frac{\varphi}{3} \frac{1 + 6 e^3}{1 + e^2}\right]
\end{equation*}
as explained at 
\url{https://ebruneton.github.io/black_hole_shader/black_hole/functions.glsl.html}.

\section{Disc particles}\label{sec:discparticles}

The orbit of a point particle in the accretion disc is given by
\begin{equation*}
u = u_1 + (u_2 - u_1) \mathrm{sn}^2 
    \left(\frac{\phi}{2} \sqrt{u_3 - u_1}, \kappa\right), \quad
\kappa = \sqrt{\frac{u_2 - u_1}{u_3 - u_1}}
\end{equation*}
where $\mathrm{sn}$ is the Jacobi-sn function, $u_1 \le u_2 \le 1/3$, and 
$u_3 = 1 - u_1 - u_2$ \cite{darwin1959}. For quasi-circular orbits this can be 
approximated with
\begin{equation}
\begin{split}
u(t) &\approx 
    u_1 + (u_2 - u_1) \sin^2\left(\frac{\pi}{4K}\phi(t)\sqrt{u_3 - u_1}\right)\\
\phi(t) &\approx
    \sqrt{\frac{\bar{u}^3}{2}}\ t + \phi_0, \quad  
    K = \int_0^1 \frac{\diff x}{\sqrt{(1 - x^2)(1 - \kappa^2x^2)}}
\end{split}\label{eq:approx-disc-orbit}
\end{equation}
where $\bar{u} = (u_1 + u_2) / 2$ since, for circular orbits, 
\eqref{eq:k_prime} gives $\diff\phi / \diff t = \sqrt{u^3 / 2}$. For a linear 
particle parameterized by $a \in [0,2\pi[$, the position $u_a(t), \phi_a(t)$ of 
a point $a$ is obtained by replacing $\phi(t)$ with $\phi_a(t) = a + \phi(t)$ 
in \eqref{eq:approx-disc-orbit}. Thus, given a ray hit point $h^t, h^r, 
h^\phi$, we compute the parameter $a$ of the "nearest" particle point with $a = 
h^\phi - \phi(h^t) \mod 2\pi$. We then compute the "distance" between $h$ and 
the linear particle center (at $a = \pi$) with $d^2 = (a / \pi - 1)^2 + (h^r - 
1 / u_a(h^t))^2$. We finally compute the particle density at $h$ with a 
smoothly decreasing function of $d$.

\section{Camera orbit}\label{sec:cameraorbit}

\paragraph*{Position} The camera position is specified by its polar coordinates 
$(r,\psi)$ in an orbital plane with inclination $\chi$ (see 
Fig.~\ref{fig:orbit}). In Schwarzschild coordinates adapted to this orbital 
plane the camera 4-velocity is $\vec{k}_c = [\frac{\diff t}{\diff \tau}, 
\frac{\diff r}{\diff \tau}, 0, \frac{\diff \psi}{\diff \tau}]^\top = 
[\frac{e}{1 - u}, \frac{\diff r}{\diff \tau}, 0, l u^2]^\top$, where $e$ and 
$l$ are two constants of motion. Substituting this in \eqref{eq:metric} gives
\begin{equation*}
\left(\frac{\diff r}{\diff \tau}\right)^2 = e^2 + l^2 u^3 - l^2 u^2 + u - 1
\Rightarrow
\frac{\diff^2 r}{\diff \tau^2} = \frac{2 l^2 u^3 - 3 l^2 u^4 - u^2}{2}
\end{equation*}
We use these relations to update the coordinates $(t, r, \psi)$ at each proper 
time step $\diff \tau$. The corresponding Cartesian coordinates are
\begin{equation*}
r
\begin{bmatrix}
\cos\chi \cos\psi \\
\sin\psi\\
\sin\chi \cos\psi
\end{bmatrix}
=
p^r
\begin{bmatrix}
\sin p^\theta \cos p^\phi\\
\sin p^\theta \sin p^\phi\\
\cos p^\theta
\end{bmatrix}
\end{equation*}
from which we deduce the Schwarzschild coordinates $p^t = t $, $p^r = r$,
$p^\theta = \arccos(\cos\psi \sin\chi)$ and $p^\phi = \arctan2(\sin\psi, 
\cos\chi \cos\psi)$.

The above relations require the constants of motion $e$ and $l$. We compute 
them from the initial position, direction and speed, noted $r_0 = 1 / u_0$, 
$\delta_0$, and $v_0$ (see Fig.~\ref{fig:orbit}). We get $e^2 = (1 - u_0) / (1 
- v_0^2)$ from the Lorentz factor $\gamma = g(\vec{k}_c, \vec{k}_s) = 
e / \sqrt{1 - u} = 1 / \sqrt{1 - v^2}$, where $\vec{k}_s = [1 / \sqrt{1 
- u}, 0, 0, 0]^\top$ is the 4-velocity of a static observer. Finally, using 
$\tan\delta = r \diff\psi / \diff r$ and the above equations, we get $l^2 = 
(e^2 + u_0 - 1) / (u_0^2 (1 - u_0 + \cot^2\delta_0))$.

\paragraph*{Lorentz transform} We compute the Lorentz transform $\Lambda$ from 
the static observer basis $\vec{e}_t, \vec{e}_r, \vec{e}_\theta, 
\vec{e}_\phi$ to the camera basis $\vec{e}_\tau, \vec{e}_w, \vec{e}_h, 
\vec{e}_d$ by using the intermediate orthonormal basis $\vec{e}_t, \vec{e}_r, 
\vec{e}_\chi, \vec{e}_\psi$ where $\mathbf{e}_\chi$ is the orbital plane's 
normal (see Fig.~\ref{fig:orbit}), as follows.

Let ${R_k}^j$ be the rotation matrix from $\vec{e}_t, \vec{e}_r, 
\vec{e}_\theta, \vec{e}_\phi$ to $\vec{e}_t, \vec{e}_r, \vec{e}_\chi, 
\vec{e}_\psi$: $\vec{e}_k = {R_k}^j \vec{e}_j$, $k \in \{t, r, \chi, \psi\}$, 
$j \in \{t, r, \theta, \phi\}$. Its lower right block is
\begin{equation*}
\begin{bmatrix}
\mathbf{e}_\chi \cdot \mathbf{e}_\theta & 
    \mathbf{e}_\chi \cdot \mathbf{e}_\phi \\
-\mathbf{e}_\chi \cdot \mathbf{e}_\phi &
     \mathbf{e}_\chi \cdot \mathbf{e}_\theta
\end{bmatrix}\ \mathrm{with}
\begin{array}{l}
\mathbf{e}_\chi \cdot \mathbf{e}_\theta = 
    \sin\chi \cos p^\theta \cos p^\phi + \cos\chi \sin p^\theta \\
\mathbf{e}_\chi \cdot \mathbf{e}_\phi = \sin\chi \sin p^\phi 
\end{array}
\end{equation*}

In the $\vec{e}_t, \vec{e}_r, \vec{e}_\chi, \vec{e}_\psi$ basis the camera 
4-velocity and speed are:
\begin{equation*}
\begin{split}
\vec{k}_c &= \left[ \sqrt{1 - u} \frac{\diff t}{\diff \tau}, \frac{1}{\sqrt{1 - 
u}} \frac{\diff r}{\diff \tau}, 0, \frac{1}{u} \frac{\diff \psi}{\diff \tau} 
\right]^\top \\
\mathbf{v} &= \left[\frac{1}{1 - u} \frac{\diff r}{\diff \tau} / \frac{\diff 
t}{\diff \tau}, 0, \frac{1}{u \sqrt{1 - u}} \frac{\diff \psi}{\diff \tau} / 
\frac{\diff t}{\diff \tau} \right]^\top
\end{split}
\end{equation*}

A reference frame for the camera is thus $\vec{e}_{k'} \triangleq 
{B(\mathbf{v})_{k'}}^k \vec{e}_k$, $k' \in \{t', r', \chi', \psi'\}$, where 
$B(\mathbf{v})$ is a Lorentz boost: ${B(\mathbf{v})_{k'}}^k = 
{B(-\mathbf{v})^{k'}}_k$, $\diff p^{k'} = {B(\mathbf{v})^{k'}}_k \diff 
p^k$~\cite{weinberg1972}.

Finally, let ${O_i\,}^{k'}$ be a user specified rotation matrix. We then 
compute $\Lambda$ with ${\Lambda_i}^j = {O_i\,}^{k'} {B(\mathbf{v})_{k'}}^k 
{R_k}^j$. Note that this procedure assumes that the camera orientation is 
actively controlled, {\em i.e.} is not freely evolving as a gyroscope would be.

\bibliographystyle{alpha}  
\bibliography{paper}

\end{document}